# Emergence of extended Newtonian gravity from thermodynamics

Peter Ván [a-c], Sumiyoshi Abe [d-g]

[a] *Department of Theoretical Physics, Wigner Research Centre for Physics,
Konkoly-Thege Miklós út 29-33, 1121 Budapest, Hungary*

[b] *Department of Energy Engineering, Faculty of Mechanical Engineering, Budapest
University of Technology and Economics, Müegyetem rkp. 3, 1111 Budapest, Hungary*

[c] *Montavid Thermodynamic Research Group, Igmándi út 26, 1112 Budapest, Hungary*

[d] *Department of Physics, College of Information Science and Engineering,
Huaqiao University, Xiamen 361021, China*

[e] *Institute of Physics, Kazan Federal University, Kazan 420008, Russia*

[f] *Department of Natural and Mathematical Sciences, Turin Polytechnic University
in Tashkent, Tashkent 100095, Uzbekistan*

[g] *ESIEA, 9 Rue Vesale, Paris 75005, France*

**Abstract**   Discovery of a novel thermodynamic aspect of nonrelativistic gravity is reported. Here, initially, an unspecified scalar field potential is considered and treated not as an externally applied field but as a thermodynamic variable on an equal footing with the fluid variables. It is shown that the second law of thermodynamics imposes a stringent constraint on the field, and, quite remarkably, the allowable field turns out to be only of gravity. The resulting field equation for the gravitational potential derived from the analysis of the entropy production rate contains a dissipative term due to irreversibility. It is found that the system relaxes to the conventional theory of Newtonian gravity up to a certain spatial scale, whereas on the larger scale there emerges non-Newtonian gravity described by a nonlinear field equation containing a single coefficient. A comment is made on an estimation of the coefficient that has its origin in the thermodynamic property of the system.

## 1. Introduction

Both the laws of thermodynamics and gravity have universalities of the high level in the sense that they are independent of the details of structures and compositions of materials. This fact leads to a natural question whether these universalities are interrelated to each other. Recent developments in understanding gravity as an emergent phenomenon suggest that the answer to this question may be affirmative. Here, "emergent" means that there exists a hierarchical structure, in which gravitational dynamics is built upon thermodynamics describing the macroscopic properties of constituents governed by microscopic dynamics. In this way, the universality of gravity can be viewed in terms of that of thermodynamics. The investigations along this line have, in fact, been performed in the literature [1-5]. There, the Einstein equation combined with the holographic principle [6] has been reinterpreted first as the equation of state in equilibrium thermodynamics, an extended version of general relativity has been formulated with use of nonequilibrium thermodynamics, and proposed has been a remarkable viewpoint, which furthermore claims that gravity may be an entropic force, not fundamental.

Our purpose in the present work is to reveal a novel thermodynamic aspect of gravity in the nonrelativistic regime. Our discussion is based solely on irreversible thermodynamics, and any additional requirement such as the holographic principle is not made. Initially, we introduce a scalar field potential and treat it as a thermodynamic variable on an equal footing with the density and internal energy of the fluid coupled to the field. At this level, the physical meaning of the field is not specified yet. Then, we show that the second law imposes a stringent constraint on the form of the field energy.

We find that, quite remarkably, *the field, which can be treated as a thermodynamic variable consistently with the second law of thermodynamics, is only of gravity*. We derive a dissipative field equation from the condition on the entropy production rate. We find that the system relaxes to the conventional theory of Newtonian gravity up to a certain spatial scale, whereas, on a larger scale, there emerges non-Newtonian gravity described by the nonlinear field equation that possesses the logarithmic potential as an exact solution.

This paper is organized as follows. In Section 2, the main results of the present work are straightforwardly summarized. Both the field equation undergoing the relaxation to Newtonian theory and the one for emergent non-Newtonian gravity are presented. In Section 3, a thermodynamic analysis leading to these results is performed, in detail. It is shown that the one and only scalar field, which can consistently be treated as a thermodynamic variable, is of gravity. Finally, Section 4 is devoted to concluding remarks, in which a comment is also made on estimation of a single coefficient characterizing the non-Newtonian gravity.

**2. Relaxation to Newtonian gravity and emergence of its modification
from thermodynamics**

In thermodynamics, fields acting on materials are usually treated as external ones [7,8] and therefore are free from the constraints by the second law. In contrast, here we treat a field potential $\phi(\mathbf{r},t)$ as a thermodynamic variable. As will be shown in Section 3, *the whole framework is consistent with the second law if and only if the field energy is* *negative*. This implies that $\phi(\mathbf{r},t)$ must be the gravitational potential. We shall see that up to a certain spatial scale, the inequality describing the entropy production rate in a local equilibrium state gives rise to the following dissipative field equation:

$$\frac{\partial \phi}{\partial t} = \frac{l^2}{\tau}\left(\nabla^2\phi - 4\pi G\rho\right), \tag{1}$$

where $G$, $\rho(\mathbf{r},t)$, $\tau$ and $l$ are Newton's gravitational constant, the mass density of a fluid, the relaxation time and a macroscopic spatial scale of variations of the whole system, respectively. Time dependence of the potential has its origin in the relaxation process of the whole system. This is a diffusion-type equation with a source and describes a relaxation process to conventional Newtonian gravity characterized by the Poisson equation

$$\nabla^2\phi = 4\pi G\rho \tag{2}$$

in the limit of relaxation, where the spatial scale $l$ disappears. We note that Newtonian gravity is stable in the sense that expansion around the time-independent potential $\phi_0(\mathbf{r})$ with the static mass density, $\phi_0(\mathbf{r}) + \delta\phi(\mathbf{r},t)$, is found to yield the diffusion equation for the perturbation $\delta\phi(\mathbf{r},t)$. The "scale" mentioned above may be, for example, a galaxy scale.

On the other hand, we will find that, presumably on a larger scale such as the galaxy scale, the gravitational potential is found to relax to the solution of the following nonlinear field equation:

$$\nabla^2 \phi = K(\nabla \phi)^2, \tag{3}$$

where *K* is a characteristic coefficient. In contrast to Eq. (2) yielding the familiar $1/r$ solution, this field equation possesses the logarithmic potential (see Section 4). Thus, we see emergence of non-Newtonian gravity from thermodynamics. This highlights the prediction of our theory to be explained in the next section.

## 3. Gravitational potential as thermodynamic variable and derivations of main results

Now, we present the detailed derivations of the main results announced in Section 2. We base our argument purely on standard irreversible thermodynamics and do not introduce any additional hypothesis. What is peculiar here is that the potential of the force field is treated as a thermodynamics variable, not an external one.

Let us recall the Euler relation [9]

$$U = TS - pV + \mu N, \tag{4}$$

where, *S*, *T*, *U*, *V*, *p*, *N* and $\mu$ are the entropy, temperature, internal energy, volume, pressure, number of particles and chemical potential, respectively. A matter considered here is a fluid that is assumed to be composed of *n* different species of particles with masses $m_1$, $m_2$, ..., $m_n$. The average mass *m* is $m = (m_1 N_1 + m_2 N_2 + \cdots + m_n N_n)/N$, where $N = N_1 + N_2 + \cdots + N_n$. The mass density $\rho$ of the fluid is given by $\rho = mN/V$. It is convenient to introduce the so-called specific variables defined by $s = S/(mN) = S/(\rho V)$, $u = U/(mN) = U/(\rho V)$ and $v = V/(mN) = 1/\rho$. In terms of these, Eq. (4) is rewritten as follows:

$$u = Ts - pv + \tilde{\mu}, \tag{5}$$

where $\tilde{\mu} = \mu/m$.

Our starting point is to include the densities of the field energy and interaction energy

$$u_{\text{field}} + u_{\text{int}} = \frac{1}{2g}(\nabla\phi)^2 - \rho\phi, \tag{6}$$

in the total energy density. Here, $g$ is a coupling constant to be determined later. $\rho$ in the interaction term plays the role of an external source for the field. At this stage, $\phi$ is not specified yet, and therefore the sign change in the interaction energy is realized by redefinition of the field as $\phi \to -\phi$. Likewise, a further redefinition $\phi \to |g|^{1/2}\phi$ changes the form in Eq. (6) in an obvious way. We emphasize that $\phi$ itself is not the potential of an emergent field but fundamentally a thermodynamic variable. Thus, the total internal energy density reads

$$u = u_{\text{fluid}} + \frac{1}{2g}\left[(\nabla\phi)^2 - 2g\rho\phi\right], \tag{7}$$

where $u_{\text{fluid}}$ is the energy density of the fluid. We note that, at this point, spacetime coordinate dependence can be introduced into the variables in connection to the spatial scale of coarse graining, through which the description of the system shifts to the theory of continuum, where the temperature and pressure become spatiotemporally local. Therefore, we have from Eq. (5) the following relation expressed in terms of the specific variables:

$$T s - p v + \tilde{\mu} = u_{\text{fluid}} + \frac{1}{2 g \rho}\left[(\nabla \phi)^2 - 2 g \rho \phi\right], \tag{8}$$

provided that we are using the same notation for the field potential, although it is the coarse-grained field of the original one in Eq. (6). In this way, the field is treated as a thermodynamic variable.

On the other hand, the Gibbs relation states that

$$T d s - p d v = d u_{\text{fluid}} + \frac{1}{g} d \left[\frac{(\nabla \phi)^2}{2 \rho} - g \phi\right]. \tag{9}$$

The chemical potential does not contribute since we are working with the specific variables, that is, $dN$ being eliminated. The specific entropy is a function of $u_{\text{fluid}}$, $v(=1/\rho)$, $\phi$ and $\nabla \phi$. We reemphasize that thus the field potential is treated here as a thermodynamic variable. From Eq. (9), we obtain the following set of relations:

$$\frac{\partial s}{\partial u_{\text{fluid}}} = \frac{1}{T}, \tag{10}$$

$$\frac{\partial s}{\partial v} = \frac{p}{T} + \frac{(\nabla \phi)^2}{2gT}, \tag{11}$$

$$\frac{\partial s}{\partial \phi} = -\frac{1}{T}, \tag{12}$$

$$\frac{\partial s}{\partial (\nabla \phi)} = \frac{1}{g \rho T} \nabla \phi. \tag{13}$$

The sign on the right-hand side in Eq. (12) indicates that the value of $\phi$ at its elementary-field level is negative. Then, Eqs. (10) and (12) imply that the fluid energy and the potential are locally in equilibrium in conformity with the hydrodynamic limit.

The second law of thermodynamics is written in the laboratory frame as follows:

$$\frac{\partial}{\partial t}(\rho s) + \nabla \cdot (\rho s \mathbf{v} + \mathbf{j}) \geq 0, \tag{14}$$

where $\rho s$ is the entropy density, $\mathbf{v}$ the velocity field of the fluid responsible for the convective part of the entropy current density and $\mathbf{j}$ the conductive entropy flux.

Fluid dynamics is described by a set of equations of the conservation laws and balance. The mass conservation is given by the continuity equation

$$\frac{\partial \rho}{\partial t} + \nabla \cdot (\rho \mathbf{v}) = 0. \tag{15}$$

Then, the momentum conservation law reads

$$\frac{\partial}{\partial t}(\rho \mathbf{v}) + \nabla \cdot \left( \rho \mathbf{v} \otimes \mathbf{v} + \ddot{P} \right) = \mathbf{0}. \tag{16}$$

In the index notation, the second term on the left-hand side denotes $\left[\nabla \cdot \left(\vec{\vec{P}} + \rho \mathbf{v} \otimes \mathbf{v}\right)\right]^i = \partial_j \left(P^{ij} + \rho v^i v^j\right)$, where $\vec{\vec{P}}$ is the symmetric pressure tensor containing the field contribution, $\partial_j \equiv \partial / \partial x^j$ and Einstein's summation convention is understood for the repeated indices. And, finally the balance equation of the total internal energy is

$$\frac{\partial}{\partial t}\left(\rho u_{\text{fluid}}\right) + \nabla \cdot \left(\rho u_{\text{fluid}} \mathbf{v} + \mathbf{q}\right) = -\vec{\vec{P}} : (\mathbf{v}\vec{\nabla}), \tag{17}$$

where $\mathbf{q}$ is the heat flux containing the field potential and the following notation is used for the quantity on the right-hand side: $\vec{\vec{A}} : \vec{\vec{B}} \equiv A^{ij} B_{ji}$, that is, $\vec{\vec{P}} : (\mathbf{v}\vec{\nabla}) = P^{ij}(\partial_i v_j)$.

Using Eq. (15), we rewrite Eq. (14) as follows:

$$\rho\left[\frac{\partial s}{\partial t} + (\mathbf{v} \cdot \nabla)s\right] + \nabla \cdot \mathbf{j} \geq 0. \tag{18}$$

From Eqs. (10)-(13), we have

$$\frac{\partial s}{\partial t} = \frac{\partial s}{\partial u_{\text{fluid}}} \frac{\partial u_{\text{fluid}}}{\partial t} + \frac{\partial s}{\partial v}\frac{\partial v}{\partial t} + \frac{\partial s}{\partial \phi}\frac{\partial \phi}{\partial t} + \frac{\partial s}{\partial (\nabla\phi)} \cdot \frac{\partial (\nabla\phi)}{\partial t}$$

$$= \frac{1}{T}\frac{\partial u_{\text{fluid}}}{\partial t} + \left[\frac{p}{T} + \frac{1}{2gT}(\nabla\phi)^2\right]\frac{\partial}{\partial t}\left(\frac{1}{\rho}\right) - \frac{1}{T}\frac{\partial \phi}{\partial t} + \frac{\nabla\phi}{g\rho T} \cdot \frac{\partial (\nabla\phi)}{\partial t}, \tag{19}$$

$$(\mathbf{v}\cdot\nabla)s = \mathbf{v}\cdot\left[\frac{\partial s}{\partial u_{fluid}}\nabla u_{fluid} + \frac{\partial s}{\partial v}\nabla v + \frac{\partial s}{\partial \phi}\nabla \phi + \frac{\partial s}{\partial(\partial_i \phi)}\nabla(\partial_i \phi)\right]$$

$$= \mathbf{v}\cdot\left\{\frac{1}{T}\nabla u_{fluid} + \left[\frac{p}{T} + \frac{1}{2gT}(\nabla\phi)^2\right]\nabla\left(\frac{1}{\rho}\right) - \frac{1}{T}\nabla\phi + \frac{\partial^i \phi}{g\rho T}\nabla(\partial_i \phi)\right\}. \quad (20)$$

Substituting Eqs. (19) and (20) into Eq. (18) and using Eqs. (15)-(17), we obtain

$$\nabla\cdot\left(\mathbf{j} - \frac{\mathbf{Q}}{T}\right) + \mathbf{Q}\cdot\nabla\left(\frac{1}{T}\right) - \frac{1}{T}\ddot{\Pi}:(\mathbf{v}\bar{\nabla}) - \frac{1}{gT}\left[\frac{\partial\phi}{\partial t} + (\mathbf{v}\cdot\nabla)\phi\right]\left(\nabla^2\phi + g\rho\right) \geq 0, \quad (21)$$

where $\mathbf{Q}$ and $\left(\ddot{\Pi}\right)^{ij} = \Pi^{ij}$ are given by

$$\mathbf{Q} = \mathbf{q} - \frac{1}{g}\left[\frac{\partial\phi}{\partial t} + (\mathbf{v}\cdot\nabla)\phi\right](\nabla\phi), \quad (22)$$

$$\Pi^{ij} = P^{ij} - p\delta^{ij} - \frac{1}{2g}\left[(\nabla\phi)^2\delta^{ij} - 2(\partial^i\phi)(\partial^j\phi)\right], \quad (23)$$

respectively. Since the pressure tensor is symmetric, so is $\Pi^{ij}$. It is convenient to decompose it into the traceless and isotropic parts: $\Pi^{ij} = \Pi^{(TL)ij} + \Pi^{(ISO)ij}$, where $\Pi^{(TL)ij} = \left(\Pi^{ij} - \Pi\delta^{ij}/3\right)$, $\Pi^{(ISO)ij} = \Pi\delta^{ij}/3$ with $\Pi \equiv \Pi^i_{\ i} = P^i_{\ i} - 3p - (\nabla\phi)^2/(2g)$.

Then, the factor in the third term on the left-hand side of Eq. (21) is rewritten as follows:

$$(\mathbf{v}\bar{\nabla}):\ddot{\Pi} = \frac{1}{2}\ddot{W}^{(TL)}:\ddot{\Pi}^{(TL)} + \frac{1}{3}(\nabla\cdot\mathbf{v})\Pi, \quad (24)$$

where

$$\left(\vec{\vec{W}}^{(TL)}\right)_{ij} = W_{ij}^{(TL)} = \partial_i v_j + \partial_j v_i - \frac{2}{3}(\nabla \cdot \mathbf{v})\delta_{ij}. \tag{25}$$

With this expression, we further rewrite Eq. (21) as

$$\nabla \cdot \left(\mathbf{j} - \frac{\mathbf{Q}}{T}\right) + \mathbf{Q} \cdot \nabla\left(\frac{1}{T}\right) - \frac{1}{T}\left[\frac{1}{2}\vec{\vec{W}}^{(TL)} : \vec{\vec{\Pi}}^{(TL)} + \frac{1}{3}(\nabla \cdot \mathbf{v})\Pi\right]$$
$$- \frac{1}{gT}\left[\frac{\partial \phi}{\partial t} + (\mathbf{v} \cdot \nabla)\phi\right]\left(\nabla^2 \phi + g\rho\right) \geq 0. \tag{26}$$

This is the form of the second law of thermodynamics for the coupled fluid-field system.

To solve Eq. (26) in the linear regime, first of all, we impose the familiar condition

$$\mathbf{j} = \frac{\mathbf{Q}}{T}, \tag{27}$$

which shows how the total heat flux including the field potential in Eq. (22) is related to the conductive entropy flux. In what follows, we consider an isotropic fluid for the sake of simplicity. Then, the flux $\mathbf{Q}$ should also be linked with the temperature gradient:

$$\mathbf{Q} = -\lambda \nabla T, \tag{28}$$

where $\lambda$ is the scalar Fourier heat conduction coefficient. The shear viscosity $\eta$ is introduced through the relation

$$\ddot{\Pi}^{(TL)} = -\eta\, \ddot{W}^{(TL)}. \tag{29}$$

Equations (27)-(29) enable us to further rewrite Eq. (26) as follows:

$$\frac{\lambda}{T^2}(\nabla T)^2 + \frac{\eta}{2T}\ddot{W}^{(TL)}:\ddot{W}^{(TL)} - \frac{1}{3T}(\nabla \cdot \mathbf{v})\Pi$$
$$-\frac{1}{gT}\left(\nabla^2 \phi + g\rho\right)\left[\frac{\partial \phi}{\partial t} + (\mathbf{v}\cdot\nabla)\phi\right] \geq 0. \tag{30}$$

The third and fourth terms on the left-hand side consist of the products of the scalar quantities. Therefore, we introduce the following linear relation [7,8]:

$$\begin{pmatrix} [\partial \phi/\partial t + (\mathbf{v}\cdot\nabla)\phi]/g \\ \Pi/3 \end{pmatrix} = \begin{pmatrix} L_{11} & L_{12} \\ L_{21} & L_{22} \end{pmatrix} \begin{pmatrix} -(\nabla^2 \phi + g\rho) \\ -(\nabla \cdot \mathbf{v}) \end{pmatrix}, \tag{31}$$

where the kinetic coefficients $L_{ij}$'s depend on not only the local temperature but also *spatial scales*, and satisfy the Onsager-Casimir reciprocity relation

$$L_{21} = -L_{12}, \tag{32}$$

which has its origin in microscopic reversibility. Here, one should recall that the antisymmetry in Eq. (32) is due to the evenness and oddness under time reversal of the components of the two columns appearing in Eq. (31). Accordingly, Eq. (30) is further rewritten as follows:

$$\frac{\lambda}{T^2}(\nabla T)^2 + \frac{\eta}{2T}\ddot{W}^{(TL)}:\ddot{W}^{(TL)} + \frac{L_{22}}{T}(\nabla\cdot\mathbf{v})^2 + \frac{L_{11}}{T}(\nabla^2\phi+g\rho)^2 \geq 0, \tag{33}$$

which is now manifestly satisfied, if both $L_{11}$ and $L_{22}$ are nonnegative. In what follows, we require

$$L_{11},\ L_{22} > 0, \tag{34}$$

in order to avoid the trivial case when these coefficients vanish.

Next, following the procedure in nonequilibrium thermodynamics, we eliminate $\nabla\cdot\mathbf{v}$ from Eq. (31) to have

$$\frac{\partial\phi}{\partial t}+(\mathbf{v}\cdot\nabla)\phi = -g\frac{\det(L)}{L_{22}}(\nabla^2\phi+g\rho)+g\frac{L_{12}}{L_{22}}\frac{\Pi}{3}, \tag{35}$$

where $\det(L)$ stands for the determinant of the matrix in Eq. (31) with Eq. (32), that is,

$$\det(L) = L_{11}L_{22}+(L_{12})^2\ (>0). \tag{36}$$

Substituting the expression of $\Pi$ given below Eq. (23), we obtain

$$\begin{aligned}\frac{\partial\phi}{\partial t}+(\mathbf{v}\cdot\nabla)\phi &= -g\frac{\det(L)}{L_{22}}(\nabla^2\phi+g\rho)\\ &\quad -\frac{gL_{12}}{3L_{22}}\left[\frac{1}{2g}(\nabla\phi)^2-P^i_{\ i}+3p\right],\end{aligned} \tag{37}$$

This is the general dissipative field equation that is consistent with the second law of thermodynamics. We note that the nonlinearity $(\nabla\phi)^2$ appears through $\Pi$ since the field potential contributes to the pressure tensor as in Eq. (23). In other words, the nonlinearity emerges as a "self-source" term.

Suppose there exist two largely separated spatial scales (see the discussion in Section 4). On the smaller scale, e.g., typically up to a galaxy scale in the astrophysical context, the second term on the right-hand side in Eq. (37) with the field nonlinearity, may be negligible. In fact, we have in this case the following equation:

$$\frac{\partial \phi}{\partial t} + (\mathbf{v}\cdot\nabla)\phi = -g\frac{\det(\mathrm{L})}{L_{22}}(\nabla^2\phi + g\rho), \qquad (38)$$

which describes relaxation to Newtonian gravity, as can be seen at once. This is an inhomogeneous diffusion-type equation. Therefore, in order for the solution not to be divergent in the long-time limit, $g$ has to be negative because of Eqs. (34) and (36). Accordingly, it follows from Eq. (6) that *the energy density of $\phi$ at its elementary-field level is negative, leading to a remarkable conclusion that it must be of gravity* [10-12]. Therefore, we set

$$g = -4\pi G, \qquad (39)$$

and identify $G$ with Newton's gravitational constant. At this juncture, we see a physical point behind the presumed existence of spatial-scale separation mentioned above. At least up to the solar scale, there are no *a priori* reasons to discard Kepler's laws and thus Newtonian gravity remains unchanged. Furthermore, we have the constant $l$ with the dimension of length and the relaxation time $\tau$ as follows:

$$4\pi G \frac{\det(\mathrm{L})}{L_{22}} = \frac{l^2}{\tau}. \tag{40}$$

In the case when the relaxation time is much shorter than the time scale of the fluid convection, we obtain Eq. (1).

It is reemphasized that *the field can be treated as a thermodynamic variable consistently with the second law of thermodynamics if and only if it is of gravy.*

Now, suppose that the two terms on the right-hand side in Eq. (37) become comparable on a larger spatial scale. Then, we have

$$\nabla^2 \phi - \frac{L_{12}}{24\pi G \det(\mathrm{L})}(\nabla\phi)^2 - 4\pi G \rho - \frac{L_{12}}{3\det(\mathrm{L})}(P^i{}_i - 3p) = 0, \tag{41}$$

in the limit of relaxation.

Let us consider the case of the vacuum, where both $\rho$ and $p$ vanish. It may also be natural to consider the isotropic pressure tensor: $P^{ij} = P\delta^{ij}$. In this case,

$$P = \mathrm{const}, \tag{42}$$

as can be seen from Eq. (16). Such *P* would be regarded as a nonrelativistic analog of the cosmological constant that is assumed to be negligibly small. Then, we reach Eq. (3) with the coefficient

$$K = \frac{L_{12}}{24\pi G \det(\mathrm{L})}, \tag{43}$$

which characterizes the emergent non-Newtonian gravity that is to be discussed in the next section.

**4. Concluding remarks: Comment on non-Newtonian gravitational coefficient**

We have examined if a scalar field potential coupled to the fluid can be a thermodynamic variable consistently with the second law. We have found that the answer to this question is affirmative if and only if the field is gravitational. We have shown that Newtonian gravity appears in the limit of relaxation, up to a certain scale, whereas on a larger scale, we have discovered that there emerges non-Newtonian gravity described by the nonlinear field equation.

The present theory of emergent non-Newtonian gravity is characterized by the factor, $K$, which is given in terms of the kinetic coefficients as in Eq. (43). Although applying our theory to astrophysical problems is not within the scope of this work, it may be still of interest to make an estimation of $K$ from the observational data. To do so, first we note that Eq. (3) yields as its analytic solution the logarithmic potential

$$\phi(r) = \frac{1}{K}\ln r, \tag{44}$$

up to an additive constant canceling the dimensionality inside the logarithm. This is the singular solution of Eq. (3) independent of a boundary condition. Also, it should be recalled that it is the field equation in the vacuum.

The potential in Eq. (44) yields the specific force

$$\mathbf{f} = -\nabla \phi = -\frac{1}{K}\frac{\mathbf{r}}{r^2},\qquad(45)$$

which is the inverse law and not the familiar inverse square law. This is an attractive force if $K$ is positive, and therefore it is natural to require $L_{12}$ to be positive. Compared with the Newtonian, this force is considered to be dominant on a larger spatial scale. From the positivity of $K$, it follows that in the large-$r$ regime the system, in fact, relaxes to the stable state described by Eq. (3).

So far, the value of the coefficient $K$ is left unspecified. Although astrophysical applications of the present theory of emergent non-Newtonian gravity are outside the scope of this paper, it may still be of interest to estimate the value of $K$ in a *model-free* way. At present, the one and only possibility of performing it is to directly employ the observational data of galaxy rotation. The force in Eq. (45) gives rise to the rotational velocity of a galaxy independent of its radius. This is concerned with the celebrated problem of dark matter. Here, we do not take dark matter into account since our purpose is to estimate the model-free value of $K$. (This, however, does not mean that we are denying dark matter models: we just wish to avoid the dark matter conundrum here since as mentioned it is outside the scope of the present work.) Then, $K$ can directly be related to the magnitude of rotational velocity outside a galaxy $V_*$ as follows: $K = 1/(V_*)^2$. From the observational data [13,14], $V_*$ for various galaxies is approximately between $10\,\text{km/s}$ and $320\,\text{km/s}$. Thus, $K$ is estimated to be between

$1.0 \times 10^{-11} \text{s}^2/\text{m}^2$ and $1.0 \times 10^{-8} \text{s}^2/\text{m}^2$.

**Acknowledgements**

P.V. has been supported by the grants, National Research, Development and Innovation Office – NKFIH 124366(124508), 123815, TUDFO/51757/2019-ITM (Thematic Excellence Program) and FIEK-16-1-2016-0007, as well as the Higher Education Excellence Program of the Ministry of Human Capacities in the frame of Nanotechnology research area of Budapest University of Technology and Economics (BME FIKP-NANO). S.A. has been supported in part by a grant from National Natural Science Foundation of China (No. 11775084), the Program of Fujian Province and by the Program of Competitive Growth of Kazan Federal University from the Ministry of Education and Science of the Russian Federation. The present work has been initiated while S.A. has stayed at the Wigner Research Centre for Physics as a Distinguished Guest Fellow of the Hungarian Academy of Sciences.